\documentclass[
   ,final            
 ,numberedheadings 
  ]
  {aipproc}

\layoutstyle{6x9}

\newcommand{\be}{\begin{equation}}
\newcommand{\ee}{\end{equation}}
\newcommand{\ba}{\begin{eqnarray}}
\newcommand{\ea}{\end{eqnarray}}

\begin{document}

\title{Loop Quantum Gravity Induced Modifications  to Particle Dynamics \footnote{Dedicated to
A. Garc\'\i a and A. Zepeda on their sixtieth birthday}}

\author{Luis F. Urrutia}{
  address={Departamento de F\'\i sica de Altas Energ\'\i as\\
  Instituto de Ciencias Nucleares\\
  Universidad Nacional Aut\'onoma de M\'exico\\
  Circuito Exterior, C. U. \\
  Apartado Postal 70-543\\
  04510 M\'exico, D.F.}
}

\begin{abstract}
The  construction of effective Hamiltonians  arising from Loop
Quantum Gravity and  incorporating Planck scale corrections to the
dynamics of photons and spin 1/2 particles is summarized.  The
imposition of strict bounds upon some parameters of the model
using already existing experimental data is also reviewed.
\end{abstract}

\maketitle

\section{Introduction}

The possibility of bringing quantum gravity induced effects to the
observational realm  has sparkled a lot of attention recently. It
is expected that the observation of high energy cosmological
particles, such as photons \cite{AC98} and neutrinos \cite{URRU1}
arising from gamma ray bursts, will provide  the appropriate arena
to test such predictions. Also, very precise experiments already
performed in atomic and nuclear physics to search for  minute
Lorentz covariance violations have been used to place strict
bounds upon such effects \cite{SUDVUUR}.

One of the leading theories providing a consistent description of
quantum gravity is  Loop Quantum Gravity (LQG) \cite{RROV}.  This
theory predicts the quantization of space in units of $\ell_P^3$,
with $\ell_P$ being the Planck length \cite{volumeop}. An
intuitive way of thinking about this is to imagine space being
described by discrete cells at very small distances $d\sim
\ell_P$, with  the standard continuous description being recovered
for large distances $d\geq \ell_P$. From the point of view of a
particle immersed in such a space, this granular structure will
act as an effective media modifying the particle propagation
properties with respect to those usually assumed in the standard
vacuum. Such granularity will induce also  minute violations of
Lorentz covariance, which have been the subject of very precise
experimental investigations \cite{HUDR}, as well as theoretical
descriptions leading to  a standard model extension which can
account for the diversity of observations that have been made
\cite{kostelecky1}. Modifications arising from LQG constitute a
specific realization of such general scheme, providing a physical
interpretation of the parameters involved.

To obtain such modifications starting from the full LQG requires a
semiclassical approximation where the particles (photons and spin
1/2 particles, for example) are treated as classical fields, while
an appropriate integration is performed upon the gravitational
degrees of freedom. In this sense, we are interested in the regime
where the matter fields are slowly varying while the gravitational
variables  are rapidly varying. The  full Hamiltonian is known in
LQG, being a  well defined regularized operator acting upon
cylindrical functions. These are functions of generalized
connections defined upon graphs $\Gamma$, characterized by a set
of vertices $\{ v \}=\{v_1,v_2, \dots, \}$ and edges $\{ e
\}=\{e^{\prime}, e^{\prime \prime}, e^{\prime \prime \prime},
\dots \} $ joining those vertices. What is missing is the strict
construction of the semiclassical  state describing the matter
field of interest, together with the corresponding large scale
continuous space-time metric (flat space in our case). With these
two ingredients one would define and calculate the semiclassical
effective Hamiltonian as the expectation value of the full LQG
Hamiltonian in the corresponding semiclassical state. A rigorous
formulation of this problem has turned out to be complicated and
is presently in the process of development \cite{aei}. Here we
take an heuristical point of view, starting from the exact
operator version of LQG and defining its action upon the
semiclassical state through some plausible requirements.

Central to our approach is Thiemann's regularization of the LQG
Hamiltonian \cite{ThiemannR}. This is based upon a triangulation
of space, adapted to the corresponding graphs which define a given
state. The regularization is provided by the volume operator, with
discrete eigenvalues arising  only from the vertices of the graph.

The paper is organized as follows \footnote{Limitations of space
prevent us  to provide a more complete list of references. We
apologize to the corresponding authors. }: section 2 contains a
very compact summary of Thiemann's regularization, exemplified in
the context of the magnetic sector of QED, together with the
heuristical scheme employed in our estimations. Section 3
summarizes the results for the case of photons and spin 1/2
particles and contains a brief discussion regarding the choice of
some relevant parameters in the model. Section 4 contains the
description of the phenomena from the point of view of the
laboratory frame attached to earth, moving at a speed $v/c\approx
10^{-3}$ with respect to the Cosmic Microwave Background frame
(CMB), yielding modifications which can be tested with already
existing experimental data.

\section{The Calculation }

\subsection{The regularized Hamiltonian operator}

The curved space magnetic contribution to the QED Hamiltonian is
\begin{equation}
H^{B} = \frac{1}{Q^2}\, \int_{\Sigma} d^3x \,
\,\frac{q_{ab}}{\sqrt{q}} \, \frac{1}{2}\,\underline B ^a
\underline B^b,
\end{equation}
where ${\underline B} ^a=\epsilon_{abc}{\underline F}_{bc}, \,\,\,
{\underline F}_{bc}=\partial_b{\underline A}_c -
\partial_c{\underline A}_b$, in standard notation. The underline
identifies the electromagnetic variables. Here $q_{ab}$ is the
three-metric, $q=\det{(q_{ab})}$ with   $a,b,c, \dots$ being
spatial indices and $Q$ is the electromagnetic coupling constant.
Thiemann's regularized expression for the above contribution is
\cite{ThiemannR}
\begin{eqnarray}
\label{REGHAM} {\hat H}^{B}
&=&\frac{1}{2\,\ell_P^4\,Q^{2}}\sum_{v\in V(\gamma
)}\,G(v)\sum_{v(\Delta )=v({\Delta'})=v}\epsilon ^{JKL}\, \epsilon
^{MNP}\times \nonumber \\
&& \times{\hat w}_{i L \Delta}\, \,\left( {\underline{h}} _{\alpha
_{JK}(\Delta )}-1\right)  \, {\hat w}_{i P \Delta'} \,\left({
\underline{h}}_{\alpha _{MN}(\Delta')}-1\right),
\end{eqnarray}
where
\begin{equation}
\label{gravv} {\hat w}_{k I \Delta}= Tr\left( \tau_k
h_{s_I(\Delta)} \left[ h^{-1}_{s_I(\Delta)},\sqrt{{\hat
V}_v}\right] \right),\nonumber
\end{equation}
with ${\hat V}_v$ being the volume operator.  The above equations
are written in terms of the holonomies $h_\gamma$ of the
corresponding connections: $A_a^i$ (gravitational) and
${\underline A}_a$ (electromagnetic) along the curves $\gamma:
s_I(\Delta), \alpha_{IJ}(\Delta)$, to be defined below. The
triangulation involved in (\ref{REGHAM}) is adapted  to the graph
$\Gamma$ corresponding to the state acted upon, in such a way that
at each vertex $v$ of $\Gamma$ and triplet of edges
$e,e^{\prime},e^{\prime\prime}$ joining the vertex, a tetrahedron
is defined with basepoint at the vertex $v(\Delta)=v$ and segments
$s_I(\Delta),\, I=1,2,3$, in the directions of $e, e^{\prime},
e^{\prime \prime}\,$ respectively.
The arcs connecting the end points of $s_I(\Delta)$ and $%
s_J(\Delta)$ are denoted $a_{IJ}(\Delta)$, so that a loop
$\alpha_{IJ}:= s_I\circ a_{IJ}\circ s_J^{-1}$ can be formed.

\subsection{The semiclassical approximation}

We think of the semiclassical configuration describing the
particular matter field (${\vec E}, {\vec B}$ in this case) plus
flat-space at large distances, as given by an ensemble of graphs
$\Gamma$, each occurring with probability $P(\Gamma)$. To each of
such graphs we associate a wave function $|\Gamma, {\cal L}, {\vec
E}, {\vec B}\, \rangle \equiv |\Gamma, S \rangle$ which is peaked
with respect to  the classical electromagnetic field configuration
together with a flat gravitational metric and a zero value for the
gravitational connection. In other words, the contribution for
each operator inside the expectation value  are  estimated  as
\cite{AMUFOT, AMUNU}
\begin{eqnarray}
\langle\Gamma, {\cal L}, \underline{{\vec{E}}},
\underline{{\vec{B}}}|\, ...{\hat q}_{ab}...\,|\Gamma, {\cal L},
\underline{{\vec{E}}}, \underline{{\vec{B}}}\rangle&=& \delta_{ab}
+ O\left(\frac{\ell_P}{\cal L}\right)
\nonumber \\
\langle\Gamma, {\cal L}, \underline{{\vec{E}}},
\underline{{\vec{B}}}|\, ...{\hat A}_{ia}...\,|\Gamma, {\cal L},
\underline{{\vec{E}}}, \underline{{\vec{B}}}\rangle&=& 0\, +
\frac{1}{\cal L}\, \left(\frac{\ell_P}{{\cal L}}\right)^\Upsilon,
\label{EXPV}
\end{eqnarray}
while the  expectation values including  the electric and magnetic
operators are estimated through their corresponding classical
values ${\vec E}$ and ${\vec B}$. The parameter $\Upsilon \geq 0$
is a real number.
 Not surprisingly, the semiclassical state specifies both the classical coordinate
 and the classical momentum for each pair of canonical variables.
 The scale ${\cal L}>> \ell_P \, $ of the wave function is such that the continuous
 flat metric approximation is appropriate for distances much larger that ${\cal L}$,
 while the granular structure of spacetime   becomes relevant when probing
 distances smaller that ${\cal L}$. Such scale will have a natural realization
 according to each particular physical situation.

We summarize now the method of calculation
\cite{URRU1,AMUFOT,AMUNU, URRU2} . For each graph $\Gamma$  the
effective Hamiltonian is defined as ${\rm H}_\Gamma= \langle
\Gamma ,S | {\hat H}_{\Gamma} |\Gamma, S \rangle$. For a given
vertex, inside the expectation value, one expands each operator in
powers of the segments $s_I(\Delta)$ plus derivatives of the
matter fields operators. Schematically, in the case of
(\ref{REGHAM}) this produces
\begin{eqnarray}
{\rm H}^B_{\Gamma} =\sum_{v\in V(\Gamma )}\,\sum_{v(\Delta )=v}\,
\langle \Gamma ,S |{\hat {\underline F}}_{p_1q_1}
(v)...\partial^{a_1}... {\hat {\underline F}}_{pq}(v) {\hat
T}_{a_1}...^{pq\,p_1q_1\,...}(v, s(\Delta))|\Gamma, S\rangle.&&
\end{eqnarray}
where ${\hat T}$ contains  gravitational operators together with
contributions depending on the segments of the adapted
triangulation in the  particular graph. Next,  space  is
considered to be divided into  boxes, each centered at a fixed
point ${\vec{x}}$ and with volume ${\cal L}^{3}\approx d^{3}\,x$.
The choice of boxes is the same for all the graphs considered.
Each box contains a large number of vertices of the semiclassical
state (${\cal L}>\!>\ell _{P}$), but it is considered as
infinitesimal in the scale where the space can be regarded as
continuous. The sum over the vertices in (\ref{REGHAM}) is
subsequently split as the sum over the vertices in each box, plus
the sum over boxes. Also, one assumes that the electromagnetic
operators are slowly varying within a   box (${\cal L}<\!<\lambda
$, with $\lambda$ been the particle wavelength), in such a way
that for all the vertices inside a given  box one can write
$\langle \Gamma, S|\dots \underline{{%
\hat{F}}}_{ab}(v)\dots |\Gamma, S \rangle = \mu {\underline
F}_{ab}({\vec{x}}). $ Here ${\ {\underline F}}_{ab}$ is the
classical electromagnetic field at the center of the box and $\mu$
is a dimensionless constant which is determined in such a way that
the standard classical result in the zeroth order approximation is
recovered. Applying  the procedure just described to
(\ref{REGHAM}) leads to
\begin{eqnarray}
{\rm H}_\Gamma^{B}&=&\sum_{{\rm Box}}\,\,\underline{{F}}_{p_{1}\,q_{1}}({\vec{x}}%
)\dots \,\,\left( \partial
^{a_{1}}\dots \,\,\underline{{\ F}}_{p\,q}({\vec{x}}%
)\right) \,\,\sum_{v\in {\rm Box}}\ell _{P}^{3}
 \sum_{{v(\Delta )=v}}\,\mu ^{n+1}\times \nonumber \\
 && \times \langle \Gamma, S|\frac{1}{\ell _{P}^{3}}{\hat{T}%
}_{a_{1}\dots }{}^{\,pqp_{1}\,q_{1}\dots }(v,s(\Delta ))|\Gamma, S
\rangle, \label{HAMG}
\end{eqnarray}
where $n+1$ is the total number of factors $F_{pq}({\vec x})$ .
The expectation value of the gravitational contribution is
supposed to be a rapidly varying function inside each box.
Finally, the effective Hamiltonian is defined as an  average over
the graphs $\Gamma$, i.e. over the adapted triangulations : ${\rm
H}^B=\sum_{\Gamma} P(\Gamma)\, {\rm H}_\Gamma^B $. This
effectively amounts to average the expectation values remaining in
each box of the sum
(\ref{HAMG}). We call this average ${{T}%
}_{a_{1}\dots }{}^{\,pqp_{1}\,q_{1}\dots }({\vec x})$ and estimate
it by demanding $T$ to be constructed from the  flat space tensors
$\delta_{ab}$ and $ \epsilon_{abc}$. In this way we are  imposing
isotropy and rotational invariance on our final Hamiltonian, which
consequently describes the modified dynamics in a specific
reference frame which we take to be the CMB frame. Also, the
scalings given in (\ref{EXPV}) together with the additional
assumptions: $\langle \Gamma, S|...{\hat V}...|\Gamma, S\rangle
\longrightarrow \ell_P^3, \,\,\, s_I^a \longrightarrow \ell_P$ are
used in this estimation. After replacing the summation over boxes
by the integral over space, the resulting Hamiltonian has the
final form
\begin{equation}
{\rm H}^{B} =\int d^{3}x\ \underline{{\ F}}_{p_{1}\,q_{1}}({%
\vec{x}})\dots \,\left( \frac{{}%
}{{}}\partial ^{a_{1}}\dots \underline{{\ F}}_{pq}({\vec{x}}%
)\right) \,\,{{T}}_{a_{1}\dots
}{}^{\,pqp_{1}\,q_{1}\dots}({\vec{x}}).  \label{AMH}
\end{equation}
Since the approach presented here has made use only of the main
features that semiclassical states should have, all dimensionless
coefficients in the expectation values that contribute to
${{T}}_{a_{1}\dots }{}^{\,pqp_{1}\,q_{1}\dots}({\vec{x}})$ in
(\ref{AMH}) remain undetermined. They are subsequently denoted by
$\theta$'s and $\kappa$'s.

\section{The results}

Here we summarize the corresponding  effective Hamiltonians and
modified dispersion relations for the cases of photons and
two-component spin 1/2 particles.

\subsection{Photons}

A detailed discussion can be found in Refs. \cite{AMUFOT, URRU2}.
The effective Hamiltonian is

\ba \label{HEMFIN} {\rm H}^{EM}= \frac{1}{Q^2}\int d^3{\vec x}
\left[\left(1+ \theta_7 \,\left(\frac{\ell_P}{{\cal
L}}\right)^{2+2\Upsilon} \right)\frac{1}{2}\left(\frac{}{}
\underline{{\vec B}}^2 + \underline{{\vec E}}^2\right) + \theta_3
\, \ell_P^2 \, \left( \frac{}{}\underline{B}^a \,\nabla^2
\underline{B}_a + \underline{E}^a \,\nabla^2
\underline{E}_a\right)+
\ \right. && \nonumber \\
\left.+ \theta_2\,\ell_P^2\,{\underline E}^a
\partial_a \partial_b {\underline E}^b+
\theta_8 \ell_P \left( \frac{}{} \underline{\vec B}\cdot(\nabla
\times\underline{\vec B})+ \underline{\vec E}\cdot(\nabla
\times\underline{\vec E}) \right) + \theta_4\, {\cal L}^2 \,
\ell_P^2 \, \left(\frac{{\cal L}}{\ell_P} \right)^{2 \Upsilon}\,
\left(\frac{}{}\underline{{\vec B}}^2\right)^2 +\dots \right],&&
\ea up to order $\ell_P^2$. The corresponding dispersion relation
is
\begin{eqnarray}
\omega_{\pm}= k\left(1+\theta_7\left(\frac{\ell_P}{{\cal L}
}\right)^{2+2\Upsilon}-2\,\theta_3\,(k\ell_P)^2\pm
2\theta_8\,(k\ell_P ) \right).
\end{eqnarray}
The $\pm$ signs correspond to the different polarizations of the
photon. From the above we obtain the speed of the photon ( $
v_{\pm}(k, {\cal L})= {\partial \omega_{\pm}(k, {\cal
L})}/{\partial k}$ )
\begin{eqnarray}
v_{\pm}=1 \pm 4\, \theta_8 \,(k \ell_P)  -6\theta_3 (k
\ell_P)^2+\theta_7\,\left(k{\ell_P}\right)^{2+2\Upsilon}+ ...  .
\label{PHOTV}
\end{eqnarray}
The last expression gives $v$ expanded to leading order in
$\ell_P$, with the estimation ${\cal L}= 1/k$. To first order in
$(k \ell_P)$ we recover  the helicity dependent correction found
already in the seminal work of  Gambini and Pullin \cite{GP}. As
far as the $\Upsilon$ dependent terms we have either a quadratic
($\Upsilon=0$) or a quartic ($\Upsilon=1$) correction. The only
possibility to have a first order helicity independent correction
amounts to set $\Upsilon=-1/2$ which corresponds to that of Ellis
et. al. \cite{ELLISFOT}. However, we do not have an interpretation
for such a value of $\Upsilon$.

\subsection{Two-component spin 1/2 particles}

The details  can be found in Refs. \cite{URRU1,AMUNU}.  The
effective Hamiltonian is

\begin{eqnarray}  \label{EFFHF}
{\rm H}_{1/2} = \int d^3 x \left[ i \ \pi(\vec x) \tau^d\partial_d
\ {\hat A} \right. \xi({\vec x}) + c.c.  + \frac{i}{4\hbar}
\frac{1}{{\cal L}} \ \pi({\vec x}) \,{\hat
C}\, \xi({\vec x})\quad \qquad  &&\nonumber\\
 + \frac{m}{2 \hbar } \xi^T({\vec
x})\ (i \sigma^2) \left( \alpha + 2\hbar\,\beta \ \tau^a
\partial_a \right)\xi({\vec x})
+\left. \frac{m}{2 \hbar} \pi^T({\vec x}) \left( \alpha  +
2\hbar\,\beta \  \tau^a \partial_a \right) (i \sigma^2) \pi({\vec
x}) \right],&&
\end{eqnarray}
where
\begin{eqnarray}\label{EFFHF1}
&&{\hat A}=\left(1 + { \kappa}_{1} \left(\frac{\ell_P}{{\cal L}} \right)^{\Upsilon+1}+ { \kappa}%
_{2} \left(\frac{\ell_P}{{\cal L}} \right)^{2\Upsilon+2} + \frac{{
\kappa}_3}{2} \
\ell_P^2 \ \ \nabla^2 \right),\nonumber\\
&&{\hat C}=\hbar \ \left({ \kappa}_4 \left( \frac{\ell_P}{\cal
L}\right)^\Upsilon
+ { \kappa}_{5} \left(\frac{\ell_P}{%
{\cal L}}\right)^{2\Upsilon+1}
+ { \kappa}_{6}\left(\frac{\ell_P}{{\cal L}} \right)^{3\Upsilon+2} +\frac{%
{ \kappa}_{7}}{2} \left(\frac{\ell_P}{\cal L}\right)^{\Upsilon}\
\ell_P^2 \ \ \nabla^2\right),
\nonumber \\
&&  \alpha= \left(1 + { \kappa}_{8} \left(\frac{\ell_P}{{\cal
L}}\right)^{\Upsilon+1}\right),
 \qquad \beta=\frac{{%
 \kappa}_9}{2\hbar} \ell_P + \frac{\kappa_{11}}{2\hbar}\ell_P
 \left(\frac{\ell_P}{\cal L} \right)^{\Upsilon+1}.
\end{eqnarray}
The corresponding dispersion relation is
\begin{eqnarray}
E_\pm(p, {\cal L})=\left[ p+\frac{m^{2}}{2p}\pm \ell _{P}\left(
\frac{1}{2}m^{2}\kappa
_{9}\right) +\ell _{P}^{2}\left( -\frac{1}{2}\kappa _{3}p^{3}+\frac{1}{8}%
\left( 2\kappa _{3}+\kappa _{9}^{2}\right) m^{2}p\right) \right]\quad \qquad &&\nonumber  \\
+\left( \frac{\ell _{P}}{{\cal L}}\right) ^{\Upsilon+1 }\left[
\left( \kappa _{1}p-\frac{\Theta _{11}m^{2}}{4p}\right)
\pm \ell_P\,\left( -\kappa _{7}\frac{p^{2}}{4}+\Theta _{12}\frac{m^{2}%
}{16}\right) \right] +\left( \frac{\ell _{P}}{{\cal L}}\right)
^{2\Upsilon +2}\left( \kappa _{2}p-\frac{m^{2}}{64p}\Theta
_{22}\right),&& \label{CDR2}
\end{eqnarray}
where the new  coefficients $\Theta$ are linear combinations of
some  $\kappa$'s.  The velocity  ($ v_\pm(p, {\cal L})= {\partial
E_\pm(p, {\cal L})}/{\partial p}$)  is
\begin{eqnarray}
v_\pm(p, {\cal L})=\left[ \left( 1-\frac{m^{2}}{2p^{2}}%
\right) +\ell _{P}^{2}\left( -\frac{3}{2}\kappa
_{3}p^{2}+\frac{1}{8}\left(
2\kappa _{3}+\kappa _{9}^{2}\right) m^{2}\right) \right] \qquad \quad &&\nonumber \\
+\left( \frac{\ell _{P}}{{\cal L}}\right) ^{\Upsilon+1 }\left[
\left( \kappa _{1}+\frac{\Theta _{11}m^{2}}{4p^{2}}\right)\mp\,
\frac{\kappa _{7}}{2}(\ell_P\,p) \right]  +\left( \frac{\ell
_{P}}{{\cal L}}\right) ^{2\Upsilon +2}\left( \kappa
_{2}+\frac{m^{2}}{64p^{2}}\Theta _{22}\right),&& \label{VELG}
\end{eqnarray}
within the same approximation. Alternative results based on a
string theory inspired approach can be found in Ref.
\cite{ELLISNU}.

\subsection{The parameters $\cal{L}$ and $\Upsilon$ }

In order to produce numerical estimations of some of the effects
arising from the modifications to the dynamics previously
obtained, we must further fix the value of the scales $\cal L$ and
$\Upsilon$. Recall that ${\cal L}$ is a scale indicating the onset
distance from where the non perturbative states of the
spin-network can be approximated by the classical flat metric. The
propagating particle (photon or neutrino) is characterized by
energies which probe to distances of order $\lambda$. In order to
preserve the  description in terms of a  classical continuous
equation it is necessary that  ${\cal L}< \lambda$. Two
distinguished cases arise: (i) the mobile scale, where we take the
marginal choice ${\cal L}= \lambda$  and (ii) the universal scale,
which has been considered in Ref.\cite{AP02} in the context of the
GZK anomaly. The study of the different reactions involved
produces a preferred bound on ${\cal L}: \, 4.6\times
10^{-8}GeV^{-1}\geq{\cal L}\geq 8.3 \times 10^{-9}GeV^{-1}$. A
recent study of the gravitational Cerenkov effect together with
neutrino oscillations \cite{LAMBIASE} produces a universal scale
estimation which is consistent with the former . Bounds for
$\Upsilon$ have been estimated in Ref. \cite{AMUNU} based on the
observation that atmospheric neutrino oscillations at average
energies of the order $10^{-2}-10^2 $ GeV are dominated by the
corresponding mass differences via the oscillation length $L_m$.
This means that additional contributions to the oscillation
length, in particular the quantum gravity correction $L_{QG}$,
should satisfy $L_{QG}> L_m$. This is used to set a lower bound
upon $\Upsilon.$ Within the proposed two different ways of
estimating the scale ${\cal L}$ of the process we obtain: (i)
$\Upsilon > 0.15$ when ${\cal L}$ is considered as a mobile scale
and (ii) $1.2 <\Upsilon $ when the scale ${\cal L}$ takes the
universal value ${\cal L}\approx 10^{-8}\, GeV^{-1}$.

\section{Observational bounds using existing data}

The previously found Hamiltonians were obtained under the
assumption of flat space isotropy so that they  account for the
dynamics in a preferred reference frame. We have identified it as
the frame in which  the Cosmic Microwave Background looks
isotropic. Our velocity $\mathbf w$ with respect to that frame has
already been determined to be $w/c\approx 1.23 \times 10^{-3}$ by
COBE. Thus, in  the earth reference frame one expects the
appearance of signals indicating minute violations of space
isotropy encoded in   $\mathbf w$-dependent terms appearing in the
transformed Hamiltonian or Lagrangian \cite{SUDVUUR}. On the other
hand, many high precision experimental test of rotational
symmetry, using atomic and nuclear system, have been already
reported in the literature. Amazingly such precision is already
enough to set very stringent bounds on some of the parameters
arising from the quantum gravity corrections. In Ref.
\cite{SUDVUUR} we have considered the case of non-relativistic
Dirac particles obtaining corrections which involve the coupling
of the spin to the CMB velocity together with a quadrupolar
anisotropy of the inertial mass. The calculation was made with the
choices $\Upsilon=0$ and ${\cal L}=1/M$, where $M$ is the rest
mass of the fermion. Keeping only terms linear in $\ell_P$, the
equation of motion  arising from the two-component Hamiltonian
(\ref{EFFHF}) can be readily extended to the Dirac case as
\begin{equation}
\left( i\gamma ^{\mu }\partial _{\mu }\,+\Theta_1 m\ell
_{P}\;i\mathbf{\gamma}\cdot \nabla -\frac{K}{2}\gamma _{5}\gamma
^{0}-m\left( \alpha -i\Theta_2 \ell _{P}\,{{\Sigma}}\cdot \nabla
\right) \right) \Psi =0, \label{DIREQ}
\end{equation}
where we have used the representation in which $\gamma _{5}$ is
diagonal, the spin operator is\ $\Sigma ^{k}=(i/2)\epsilon
_{klm}\gamma ^{l}\gamma ^{m}$, $K=\Theta_4\,m^2\,\ell_P$ and
$\alpha=1+\Theta_3\,m\,\ell_P$. The normalization has been chosen
so that in the limit $(m\ell_P)\rightarrow 0$ we recover the
standard massive Dirac equation. The term $m\left(
1+{\Theta_3\;}m\ell _{P}\right) $ can be interpreted as a
renormalization of the mass whose physical value is taken to be
$M=m\left( 1+{\Theta_3\;}m\ell _{P}\right) $. After this
modification the corresponding effective Lagrangian is  \ba
L_{D}&=&\frac{1}{2}i\bar{\Psi}\gamma ^{0}\left( \partial _{0}\Psi \right) +%
\frac{1}{2}i\bar{\Psi}\left(\frac{}{}(1+\Theta_1M\ell
_{P})\gamma^k-\Theta_2\ell _{P}M\Sigma ^{k}\, \right)\, \partial
_{k}\Psi \nonumber \\
&&-\frac{1}{2}M\bar{\Psi}\Psi-
\frac{K}{4}\bar{\Psi}\gamma_5\gamma^0\Psi +{\rm h.c.},
\label{DIRLAG} \ea which describes the time evolution as seen in
the CMB frame. In order to obtain the dynamics  in the laboratory
frame we implement an observer Lorentz transformation. To this end
we rewrite (\ref{DIRLAG}) in a covariant looking form, by
introducing explicitly the CMB frame's four velocity $W^{\mu
}=\gamma (1,\,{\mathbf{w}}/c)$. In the metric with signature $-2$
the result is
\begin{eqnarray}
L_{D}&=&\frac{1}{2}i\bar{\Psi}\gamma ^{\mu }\partial _{\mu }\Psi -\frac{1}{2}M%
\bar{\Psi}\,\Psi +\frac{1}{2}i(\Theta_1 M\ell
_{P})\bar{\Psi}\gamma _{\mu }\left(
g^{\mu \nu }-W^{\mu }W^{\nu }\right) \partial _{\nu }\Psi \nonumber \\
&&+\frac{1}{4}(\Theta_2M\ell_{P})\bar{\Psi}\epsilon _{\mu \nu
\alpha \beta }W ^{\mu }\gamma ^{\nu }\gamma ^{\alpha }\partial
^{\beta }\Psi - \frac{1}{4}(\Theta_4 M\ell_P)M W_\mu
\bar{\Psi}\gamma_5\gamma^\mu\Psi +h.c. . \label{otra}
\end{eqnarray}
From Eq. (26) of \cite{Lane}  we obtain the non-relativistic limit
of the Hamiltonian corresponding to (\ref{otra}), up to first
order in $\ell_P$ and  up to order ${({\bf w})/c}^{2}$, which is
\begin{eqnarray}
\tilde{H}= \left[ Mc^2(1+\Theta_1\,
M\ell_P\,\left({\mathbf{w}}/{c} \right) ^{2}) +
\left(1+2\,\Theta_1M\ell
_{P}\left(1+\frac{5}{6}\left({\mathbf{w}}/{c} \right)
^{2}\right)\right)\left(\frac{p^{2}}{2M}+g\,\mu
\,{\bf s}%
\cdot {\bf B}\right) \right]&&\nonumber\\
 + \left(\Theta_2+\frac{1}{2}\Theta_4 \right)M\ell
_{P}\left[\left(2Mc^2
-\frac{2p^{2}}{3M}\right)\,{\bf s}%
\cdot \frac{\bf w}{c}+\frac{1}{M}\,{\bf s}\cdot {Q}_{P}\cdot
\frac{\bf w}{c}\right]
+\Theta_1M\ell _{P} %
\left[ \frac{{\bf w}\cdot {Q}_{P}\cdot {\bf w}}{Mc^2}\right],&&
\label{B6}
\end{eqnarray}
where ${\bf s}={\bf \sigma}/2$. Here we have not written the terms
linear in the momentum since they average to zero. In (\ref{B6})
$g$ is the standard gyromagnetic factor, and $Q_{P}$ is the
momentum quadrupole tensor with components
$Q_{Pij}=p_{i}p_{j}-1/3p^{2}\delta _{ij}$. The terms in the second
square bracket represent a coupling of the spin to the velocity
with respect to the ``rest'' (privileged) frame. The first one has
been measured with high accuracy in references \cite{HUDR} where
an upper bound for the coefficient has been found. The second term
is a small anisotropy contribution and can be neglected. Thus we
find the correction \be \delta H_{S}= \left(\Theta_2
+\frac{1}{2}\Theta_4\right) M\ell _{P} (2Mc^2) \left[ 1 + O\left(
\frac{p^{2}}{2M^2c^2}\right) \right] \mathbf{s}\cdot
\frac{\mathbf{w}}{c}. \ee The last term of (\ref{B6}), which
represents an anisotropy of the inertial mass, has been bounded in
Hughes-Drever like experiments. With the approximation
$Q_{P}=-5/3<p^{2}>Q/R^{2}$ for the momentum quadrupole moment,
with $Q$ being the electric quadrupole moment and $R$ the nuclear
radius, we obtain
\begin{equation}
\delta H_{Q}=-\Theta_1 M\ell _{P}\frac{5}{3}\left\langle \frac{p^{2}}{2M}%
\right\rangle \left( \frac{Q}{R^{2}}\right) \left(
\frac{w}{c}\right) ^{2}P_{2}(\cos \theta ), \label{QMM}
\end{equation}
for the quadrupole mass perturbation, where $\theta$ is the angle
between the quantization axis and $\mathbf{w}$. Using
$<p^{2}/2M>\sim 40$ MeV for the energy of a nucleon in the last
shell of a typical heavy nucleus, together with the experimental
bounds of references \cite{HUDR} we find \cite{SUDVUUR}
\begin{equation}
\mid \Theta_2+\frac{1}{2}\Theta_4\mid <2\times 10^{-9},\qquad \mid
\Theta_1\mid <3\times 10^{-5}. \label{Result2}
\end{equation}
The above  bounds  on terms that were formerly expected to be of
order unity, already call into question the scenarios inspired on
the various approaches to quantum gravity, suggesting the
existence of Lorentz violating Lagrangian corrections which are
linear in Planck's length. To this respect it is interesting to
notice that a very reasonable fit to the gamma ray spectrum beyond
de GZK cutoff has been recently made by using  dispersion
relations of higher order than linear in $\ell_P$ \cite{AP2002}.
Observational bounds upon parameters of related theories are
obtained in the works of  Ref. \cite{ADD}

\begin{theacknowledgments}
  The author wishes to thank  the organizers of the X Mexican
  School of Particles and Fields for the invitation to
  participate. Partial support from the  project DGAPA-IN-117000 is
  also acknowledged.
\end{theacknowledgments}


\begin{thebibliography}{99}

\bibitem{AC98}Amelino-Camelia, G., Ellis, J.,  Mavromatos, N.E.,
Nanopoulos, D.V. and  Sarkar, S., {\it Nature}, {\bf 393},763
(1998).

\bibitem{URRU1} Alfaro, J., Morales-T\'ecotl, H.A.,  and Urrutia,
L.F.,
{\it Phys. Rev. Letts.}, {\bf 84}, 2318 (2000).

\bibitem{SUDVUUR} Sudarsky, D., Urrutia, L.F., and  Vucetich,
H., {\it Phys. Rev. Letts.}, {\bf 89}, 231301 (2002).

\bibitem{RROV} For a recent reviews see for example: Rovelli, C., Loop quantum gravity,
{\it Livings Reviews}, {\bf 1}, 1 (1998), URL
http://www.livingreviews.org/Articles; R., and Pullin, J., {\it
Loops, Knots, Gauge Theories and Quantum Gravity}, Cambridge
University Press, Cambridge UK, 1996; Thiemann, T., gr-qc/
0110034.

\bibitem{volumeop}
 Rovelli, C., and Smolin, L., { \it Nucl. Phys.}, {\bf B442}, 593-622 (1995);
Erratum-ibid {\bf B456}, 753 (1995);  Ashtekar, A., and
Lewandowsky, J., {\it  Class. Quant. Grav.}, {\bf 14},A55-A82
(1997); { \it Adv. Theor. Math. Phys.}, {\bf 1}, 388-429 (1998);
Ashtekar, A., Corichi, A., and  Zapata, J., {\it  Class. Quant.
Grav.}, {\bf 15},2955-2972 (1998).

\bibitem{HUDR} Hughes, V.W.,  Robinson, H.G., and  Beltr\'an-L\'opez, V., {\it Phys.
Rev. Letts.}, {\bf 4}, 342 (1960); Drever,  R.W.P., {\it  Philos.
Mag.}, {\bf 6}, 683 (1961);  Prestage, J.D., { et al.}, {\it Phys.
Rev. Letts.}, {\bf 54}, 2387 (1985);  Lamoreaux, S.K., {et al.},
{\it Phys. Rev. Letts.}, {\bf 57}, 3125 (1986);  Lamoreaux, S.K.,
{ et al.}, {\it Phys. Rev.}, {\bf A 39}, 1082 (1989); Berglund, C.
J., { et al.}, {\it Phys. Rev. Lett.}, {\bf 75}, 1879 (1995);
Phillips, D. F.. { et al.}, {\it Phys. Rev. D}, {\bf 63}, 111101
(2001); Chupp, T.E. , { et al.}, {\it Phys. Rev. Letts.}, {\bf
63}, 1541 (1989); Bear, D. { et al.}, {\it Phys. Rev. Letts.},
{\bf 85}, 5038 (2000).

\bibitem{kostelecky1} Kostelecky, V.A., and  Samuel, S., {\it Phys. Rev. D}, {\bf 39},
683 (1989), {\it  Phys. Rev. D}, {\bf 40}, 1886 (1989);
Kostelecky, V.A., and  Potting, R., {\it  Nucl. Phys. B}, {\bf
359},545 (1991), {\it Phys. Lett. B}, {\bf 381}, 89
(1996); Colladay, D, and  Kostelecky, V.A., {\it Phys. Rev. D}, {\bf %
55}, 6760 (1997); {\it Phys. Rev. D}, {\bf %
58}, 116002 (1998).

\bibitem{aei} Thiemann, T.,
 {\it Class. Quant. Grav.}, {\bf 18}, 2025-2064 (2001);  Thiemann,
T., and  Winkler, O., {\it Class. Quant. Grav.}, {\bf 18},
2561-2636 (2001);  Thiemann, T., and  Winkler, O., hep-th/0005234;
hep-th/0005235; Sahlmann, V.,  Thiemann, T., and Winkler, O.,
gr-qc/0102038.

\bibitem{ThiemannR} Thiemann, T., {\it Class. Quant. Grav.}, {\bf 15},1281-1314 (1998);
 {\it Class. Quant. Grav.}, {\bf 15}, 839-873 (1998).

\bibitem{AMUFOT} Alfaro, J.,  Morales-T\'ecotl, H.A.,  and  Urrutia, L., {\it Phys. Rev. D},
{\bf 65}, 103509 (2002).

\bibitem{AMUNU} Alfaro, J.,  Morales-T\'ecotl, H.A.,  and  Urrutia, L., {\it Phys. Rev. D},
{\bf 66}, 124006 (2002).

\bibitem{URRU2}  Urrutia, L.F.,  Mod. Phys. Letts. {\bf A17},943 (2002).

\bibitem{GP} Gambini, R., and  Pullin, J., {\it Phys. Rev. D}, {\bf 59}, 124021
(1999).

\bibitem{ELLISFOT} Ellis, J.,  Mavromatos, N.E., and  Nanopoulos, D.V.,
 {\it  Gen. Rel. Grav.}, {\bf 32}, 127 (2000); {\it
Phys. Rev. D} {\bf 61}, 027503 (2000).

\bibitem{ELLISNU} Ellis, J., { et al.}, {\it Astrophysical Jour.}, {\bf 535 },
139 (2000); Ellis, J., { et al.}, {\it Gen. Rel. Grav.}, {\bf 32},
1777 (2000).

\bibitem{AP02}Alfaro, J., and Palma, G., {\it Phys. Rev. D}, {\bf 65}, 103516
(2002).

\bibitem{LAMBIASE} Lambiase, G., gr-qc/0301058.



\bibitem{Lane}  Kostelecky, V. A., and  Lane,C.D., J. Math. Phys., {\bf 40}, 6245 (1999);
 Colladay, D., and  MacDonald, P., hep-ph/0202066.

\bibitem{AP2002} Alfaro, J., and Palma, G.,  hep-th/0208193.

\bibitem{ADD}  Biller,  S.D., {\it et. al.} , {\ Phys. Rev. Lett.} {\bf 83},
2108 (1999);  L\"{a}mmerzhal, C.,  and  Bord\'e, C., in Lecture
Notes in Physics 562, Springer,  2001;  Gleiser, R.J.,  and
Kozameh, C. N.,  Phys. Rev. {\bf D64}: 083007 (2001);  Brustein,
R., Eichler, D., and  Foffa, S., Phys. Rev. {\bf D65}: 105006
(2002); Lambiase, G., Gen. Rel.Grav. {\bf 33}, 2151 (2001);
Amelino-Camelia, G., Phys. Letts. {\bf B528}, 181 (2002);
Jacobson, T.,  Liberati, S., and  Mattingly, D., hep-ph/0112207;
Konopka T., and  Major, S., New J. Phys. {\bf 4}:57 (2002);
Kostelecky, V.A., and  Lane, C.D.,  Phys. Rev. {\bf D60}, 116010
(1999).






\end{thebibliography}
\end{document}